\documentclass[cameraready]{Interspeech}
\usepackage{makecell}

\title{What Does a Pathological Speech Assessment Model Know about Acoustic Features? A Case Study on Oral and Oropharyngeal Cancer Patients}

\author[affiliation={1}, orcid=0009-0006-7047-7682]{Tuan}{Nguyen}
\author[affiliation={1}, orcid=0000-0002-0413-8950]{Corinne}{Fredouille}
\author[affiliation={3}, orcid=0000-0002-0413-8950]{Alain}{Ghio}
\author[affiliation={3}, orcid=0000-0002-7672-8589]{Muriel}{Lalain}
\author[affiliation={2,4}, orcid=0000-0003-3895-2827]{Virginie}{Woisard}

\address{
    $^1$ Avignon University, LIA, UPR 4128, Avignon, France \\
    $^2$ Hôpital Larrey, Hôpital de Toulouse, Toulouse, France \\
    $^3$ Aix Marseille University, CNRS, LPL, Aix-en-Provence, France \\
    $^4$ UT2J, Laboratoire de NeuroPsychoLinguistique, Toulouse, France
}

\email{manh-tuan.nguyen@univ-avignon.fr}


\usepackage{comment}

\begin{document}

\maketitle

\begin{abstract}
    This work investigates the interpretability of a Wav2Vec 2.0-based speech intelligibility assessment model for oral and oropharyngeal cancer patients through canonical correlation analysis.
    By measuring the correlation between the model embeddings and eGeMAPS low-level descriptors (LLDs) as an interpretable reference, we analyze how acoustic information is encoded across the model layers.
    The analysis is conducted at two levels: individual LLDs layer-wise, and group-level: prosodic, spectral, and voice quality.
    Results show that the learned representations are most strongly correlated with spectral and prosodic features, with the first MFCC coefficient yielding the highest correlations across all layers.
    At the group level, spectral and prosodic groups achieve correlations of 0.77 and 0.71 respectively, while voice quality reaches 0.65.
    Beyond model interpretability, this work also offers practical guidance on acoustic feature selection for pathological speech assessment.
\end{abstract}

\section{Introduction}
Speech processing has a long history of development, with advances spanning text-to-speech, speech recognition, and speech translation, among others, and pathological speech assessment has increasingly benefited from these advances.
Traditionally, speech and voice disorders are evaluated through human-based clinical assessment, which is inherently subjective, prone to inter-rater variability, and difficult to reproduce consistently.
These limitations have motivated a growing interest in automatic, technology-based assessment systems.
However, unlike typical speech processing, pathological speech comes from patients suffering from various medical conditions, which introduces unique challenges such as data scarcity for instance. 
More critically, due to the medical context, any technology applied to pathological speech must be interpretable: clinicians need to understand the reasoning behind system decisions, not just its outputs. 
This requirement has considerably limited the adoption of deep learning in this domain, despite its advantages. 
As a result, the pathological speech and voice community remains caught between deep learning-based approaches and more transparent, and interpretable handcrafted acoustic features \cite{FAVARO2023107559}.

A number of studies have explored pathological speech assessment using handcrafted acoustic features in recent years.
In \cite{balaguer2025}, Balaguer et al. proposed an automatic analysis combining handcrafted acoustic features with other linguistic components such as phonemic and segmental levels.
This statistical approach provides a more reliable alternative to traditional perceptual assessments for evaluating speech impairment in post-cancer patients.
Similarly, \cite{VASQUEZCORREA201821} provided different handcrafted features extracted from multiple speech dimensions, including phonation, articulation, and prosody to automatically predict the dysarthria level of Parkinson's disease patients. 
Beyond these works, handcrafted features have been widely adopted across various pathological speech conditions \cite{xue19_slate, np18_interspeech, glottal,app10196999}.
Despite a large body of work relying on handcrafted acoustic features, the choice of features varies considerably across studies, and no clear consensus or baseline feature set has been established for pathological speech assessment.
On the other hand, deep learning approaches have also gained significant attention given their success in a wide range of tasks. 
Various works have directly applied deep learning architectures to atypical speech \cite{tuanlrec,10094921, 8444654,vasquezcorrea17_interspeech}, with architectures evolving from convolutional neural networks to self-supervised learning (SSL) models, consistently achieving strong results. 
As shown in \cite{FAVARO2023107559}, DNN-based features have been shown to outperform interpretable handcrafted features in terms of performance. 
However, as discussed above, deep learning approaches lack interpretability, and therefore a trade-off between performance and interpretability remains a central challenge in this domain.

Given this trade-off, this work aims to bridge these two approaches by providing an interpretable analysis of deep learning models.
To do so, a set of handcrafted acoustic features, widely considered interpretable in the clinical community, is used as a reference to explain decisions of deep learning architectures. 
This work makes two contributions. 
First, we provide an interpretable analysis of a speech intelligibility assessment model (fine-tuned from a Wav2Vec 2.0 pre-trained for ASR) by measuring its layer-wise alignment with eGeMAPS features.
Second, we derive a practical feature importance ranking, providing the community with practical guidance on acoustic feature selection for pathological speech assessment.

\section{Methodology}
In this paper, we address the trade-off between handcrafted features and deep learning-based representations by focusing on the interpretability of the latter.
To this end, we employ Canonical Correlation Analysis (CCA) \cite{pearson1992breakthroughs} to bridge these two worlds.

CCA, first introduced by Hotelling in 1936 \cite{hotelling1936simplified}, is a statistical method that measures the linear relationships between two multivariate spaces. 
A common early application involved measuring relationships between feature sets and outputs, enabling the selection of appropriate input features for downstream tasks \cite{6854298,zhu2014multi}.
In 2017, researchers from Google \cite{svcca} extended CCA for interpretability purposes, proposing Singular Vector CCA (SVCCA), which enables the analysis of learned representations in deep neural networks.
The SVCCA method combines Singular Value Decomposition (SVD) for dimensionality reduction with CCA for measuring representational similarity, enabling practical comparison of representations from layers or models with different sizes. 
By applying SVD prior to CCA, the method removes low-variance directions that primarily introduce noise rather than meaningful signal, making it more robust for interpreting deep learning representations. 
The authors demonstrated the effectiveness of that method by analyzing how representations evolve across layers within a single model and by measuring representational similarity between different network architectures. 
Building upon this, the same authors subsequently proposed Projection-Weighted CCA (PWCCA) \cite{pwcca}, which addresses a key limitation of SVCCA. 
Unlike SVCCA, which requires manually determining a dimensionality threshold after SVD, PWCCA computes a weighted mean of correlations based on variance contribution, eliminating this requirement.

CCA and its variants have been widely used in the literature for explainability purposes \cite{9688093, 10096149, tuanslt}. 
In this work, we employ PWCCA as the main analysis tool to interpret deep learning representations by measuring their relationship with handcrafted acoustic features, which serve as an interpretable reference. 
This allows us not only to make deep learning representations more interpretable, but also to identify handcrafted features that are most correlated with deep learning representations.
Since deep learning is known to perform well, features highly correlated with its representations are strong candidates for effective yet interpretable substitutes in clinical assessment.
Thus, while promising, empirical validation of this hypothesis is left for future work.

\section{Experimental Setup}
\begin{table}[t]
  \caption{eGeMAPS LLD features grouped into clinically meaningful categories}
  \label{tab:egemaps_groups}
  \centering
  \small
  \begin{tabular}{lp{0.65\columnwidth}}
    \toprule
    \textbf{Feature Group}& \textbf{eGeMAPS LLD Features} \\
    \midrule
    \textbf{Prosodic} & Pitch (logarithmic F0) \\
     (\small 2 LLDs) & Loudness \\
    \midrule
    \textbf{Spectral} & MFCC 1--4 \\
     \small (18 LLDs) & Formants 1, 2, 3 (frequency, bandwidth, energy) \\
     & Alpha Ratio \\
     & Hammarberg Index \\
     & Spectral Slope (0--500 Hz, 500--1500 Hz) \\
    \midrule
    \textbf{Voice Quality}  & Jitter \\
     \small (5 LLDs) & Shimmer \\
     & Harmonics-to-Noise Ratio (HNR) \\
     & Harmonic difference (H1--H2, H1--A3) \\
    \bottomrule
  \end{tabular}
\end{table}
\subsection{Corpus}
This work uses the French corpus C2SI \cite{c2si}, which contains recordings from both healthy control speakers and patients diagnosed with oral and oropharyngeal cancer (OOC). 
C2SI was developed under the Carcinologic Speech Severity Index project from 2015 to 2017, initiated to address the need for objective assessment tools for evaluating speech abilities of patients with OOC. 
The corpus contains 87 patients and 40 control speakers, with 7 patients recorded twice, resulting in 134 total recording sessions.
To maintain stability of speech disorder, patients were required to complete their treatment plan at least six months before enrollment and to be in clinical remission. 
Participants were recorded on various tasks commonly used in clinical assessment, including sustained vowel, spontaneous speech, and read speech.

All participants were evaluated by a panel of 6 experts on different aspects of speech, including intelligibility. 
For the intelligibility assessment, experts listened to the audio recordings and assigned a perceptual score on a scale of 0 to 10, where 0 indicates unintelligible speech and 10 indicates fully intelligible speech. 
Then, a consensus score was derived by computing the mean opinion score among all experts. 
In this work, only the read speech task and its associated consensus intelligibility scores are considered for analysis.

\begin{figure*}[t]
  \centering
  \includegraphics[width=0.8\linewidth]{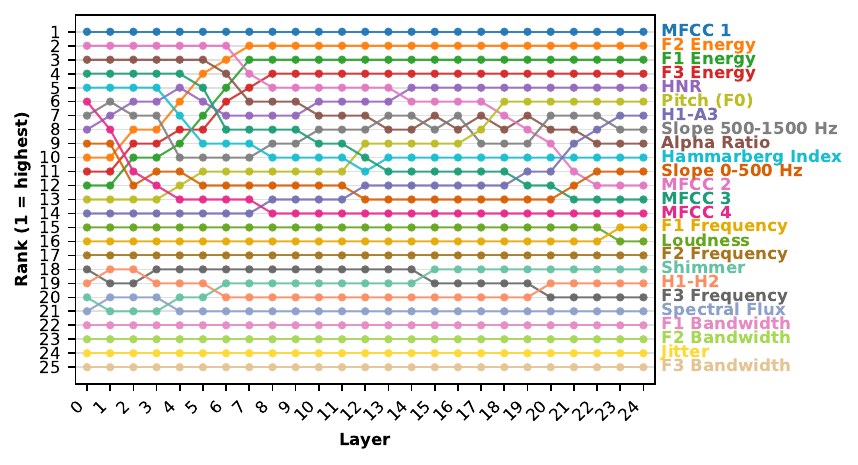}
  \caption{Layer-wise PWCCA correlation between eGeMAPS LLDs and Wav2Vec 2.0 representations (rank 1 = highest correlation).}
  \label{fig:lld_ranking}
\end{figure*}
\subsection{Speech assessment model}
With the advancement of SSL models, the current state-of-the-art approaches in pathological speech assessment typically rely on these architectures \cite{javanmardi2023wav2vec, kadiri2024investigation}. 
In the context of OOC patients, we employ the architecture proposed by Nguyen et al. \cite{tuanlrec}, which is built upon Wav2Vec 2.0 \cite{w2v2}. 
Unlike a standard SSL-based Wav2Vec 2.0, this model incorporates an intermediate fine-tuning step on an ASR task before being applied to disorder speech assessment. 
This additional step has been shown to significantly improve performance over the standard SSL-based approach. 

The architecture consists of the Wav2Vec 2.0 large configuration encoder as its core, followed by a pooling layer computing both mean and standard deviation, and two linear layers of 1024 dimensions before the output layer producing the prediction score. 
This model achieved a Mean Absolute Error (MAE) of 0.68 on intelligibility assessment within the C2SI corpus, significantly lower than comparable baselines \cite{sebastiao, sondes}. 
Given this strong performance and the central role of Wav2Vec 2.0 in the architecture, the CCA analysis is performed on the learned representations of the Wav2Vec 2.0 encoder only.
Since the encoder consists of multiple transformer layers, PWCCA is computed layer-wise, allowing us to analyze how the representations evolve across layers.

\subsection{Interpretable features}
A wide variety of handcrafted feature sets have been proposed and used across the literature, making the selection of an appropriate set a non-trivial decision. 
As this work aims not only to interpret the assessment model but also to provide guidance on which acoustic features are most relevant, selecting a well-established and widely recognized feature set is a priority. 
We therefore focus on task-specific feature sets that already aggregate multiple acoustic dimensions. 
Among the available options, we adopt the extended Geneva Minimalistic Acoustic Parameter Set (eGeMAPS) \cite{egemaps}, which offers a compact yet comprehensive alternative to larger feature sets such as ComParE \cite{compare16}, with greater interpretability and efficiency.
eGeMAPS contains 88 statistical parameters derived from 25 Low-Level Descriptors (LLDs), carefully selected based on theoretical and empirical evidence to capture relevant acoustic characteristics while maintaining a manageable feature space. 
Both sets have been broadly used in the speech processing community, spanning tasks from speech emotion recognition to pathological speech assessment \cite{xue19_slate, kshirsagar2022cross}.
Due to its efficiency and wide adoption in the literature, eGeMAPS is employed in this work as the interpretable feature set to analyze the Wav2Vec 2.0-based assessment model. 

Regarding feature type, we use LLD features rather than functional features to preserve the full temporal representation of acoustic information. 
For better clinical interpretability, we further organize the 25 LLDs into categories that correspond to speech production subsystems. Although the original eGeMAPS paper \cite{egemaps} groups these features into three engineering-oriented categories, namely frequency-related, energy/amplitude and spectral balance, this categorization does not fully align with clinical speech production frameworks. 
Therefore, based on \cite{schullerprosodic, corrales2023prosodic, VASQUEZCORREA201821}, we propose to reorganize the LLDs into three clinically motivated groups: Prosodic, Spectral, and Voice Quality. 
The details are described in Table~\ref{tab:egemaps_groups}.

\subsection{Analysis experiment}
PWCCA is used to measure the relationship between the learned representations of Wav2Vec 2.0 and eGeMAPS acoustic features.
Concretely, embeddings are extracted at the frame level from each transformer layer of the Wav2Vec 2.0-based assessment model and are correlated directly with the corresponding eGeMAPS LLD values.
Both are temporally aligned at 25ms, where the eGeMAPS LLD features are extracted using opensmile \cite{opensmile} with a frame size configured to match the Wav2Vec 2.0.
This measures how much acoustic information from eGeMAPS is encoded within the model learned representations.

We conduct two levels of analysis. 
In the \textbf{(i) individual-level analysis}, PWCCA is computed layer-wise for each individual LLD, allowing us to track how the correlation between each feature and the model representations evolves across layers. 
In the \textbf{(ii) group-level analysis}, only the last layer of Wav2Vec 2.0 is considered, as it is most directly related to the model final prediction. 
Since each group contains a different number of LLDs, the group-level correlation is computed as the mean of the individual LLD correlations within that group, ensuring a fair comparison across groups.

\section{Results}
\subsection{Individual-level analysis}
\begin{figure*}[t]
  \centering
  \includegraphics[width=0.78\linewidth]{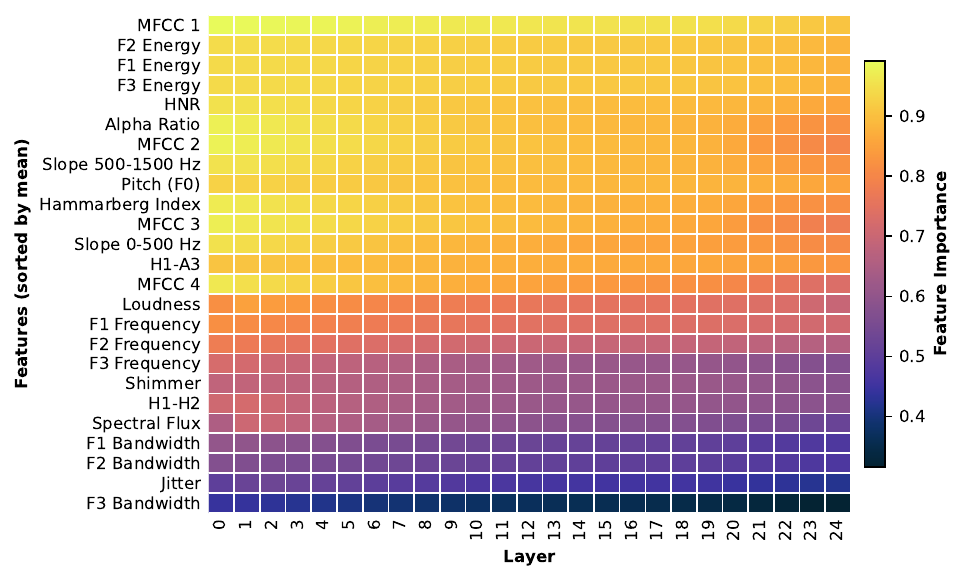}
  \caption{
 Heatmap of layer-wise PWCCA correlations between eGeMAPS LLDs and Wav2Vec 2.0 representations. Features are sorted by mean correlation across layers.}
  \label{fig:lld_heatmap}
\end{figure*}
Figure~\ref{fig:lld_ranking} illustrates the layer-wise evolution of individual eGeMAPS LLDs in terms of their PWCCA correlation ranking with Wav2Vec 2.0 representations, where rank 1 indicates the highest correlation and rank 25 the lowest. 
The ranking reveals that certain features undergo considerable changes across layers, suggesting that their contribution to the learned representations varies depending on the depth of the model.
In the early layers, MFCC features 1--4 are the most correlated with the Wav2Vec 2.0 embedding space, consistent with findings reported in \cite{9688093}.
However, as depth increases, MFCC 2--4 progressively drop in rank and appear to contribute less to the final decision of the model, giving way to formant energy-related features as well as prosodic and voice quality features such as F0 and HNR. 
MFCC 1, on the other hand, remains consistently highly ranked throughout all layers, indicating a strong and stable correlation with Wav2Vec 2.0 representations across the entire network. 
The least correlated features overall are frequency-related formant parameters and shimmer.

Figure~\ref{fig:lld_heatmap} provides a detailed view of the PWCCA correlation values between Wav2Vec 2.0 layer-wise embeddings and individual eGeMAPS LLDs, where color intensity encodes the correlation magnitude.
The visualization reveals that Wav2Vec 2.0 is strongly correlated with the majority of eGeMAPS LLDs, with MFCC 1 showing the highest correlation and F3 bandwidth the lowest.
This provides practical guidance for the community: when selecting features for pathological speech assessment, low-correlated features such as formant bandwidth, jitter, shimmer, and H1-H2 may be discarded in favor of higher-correlated features such as MFCC 1, formant energy.
It is also observed that at the final layer, where representations are most directly related to the prediction of intelligibility scores, the correlation with eGeMAPS features decreases across the board. 
This suggests that the deeper representations of Wav2Vec 2.0 may encode information beyond what eGeMAPS captures. 
This could include higher-level phonetic or linguistic cues, or additional acoustic dimensions such as glottal features, which have been shown to be relevant for speech assessment \cite{sondes2, glottal1}.
This opens future work toward extending the analysis to a wider set of features, and ultimately toward a common reference set well suited for explaining pathological speech assessment models.
Overall, the layer-wise analysis demonstrates how the decisions of Wav2Vec 2.0 relate to interpretable acoustic information referenced by eGeMAPS LLDs, while also providing the community with practical guidance on feature selection for pathological speech assessment.
\begin{figure}[t]
  \centering
  \includegraphics[width=0.77\linewidth]{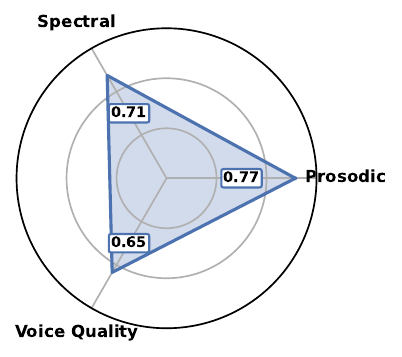}
  \caption{PWCCA correlation between the layer 24 of Wav2Vec 2.0 and feature groups: Prosodic, Spectral, and Voice Quality.}
  \label{fig:radar_groups}
\end{figure}
\subsection{Group-level analysis}
Figure~\ref{fig:radar_groups} presents the PWCCA correlation between the final layer of Wav2Vec 2.0 and the three feature groups.
The PWCCA correlation between the final layer of Wav2Vec 2.0 and the three group-level features shows that the assessment model representations are most aligned with Spectral and Prosodic features, with correlation values of 0.77 and 0.71 respectively. 
Voice Quality, while still correlated, shows a drop of 12 points compared to the Prosodic group, with a correlation of 0.65. 
Overall, these results suggest that the Wav2Vec 2.0-based assessment model relies more on spectral and prosodic information, while still incorporating voice quality cues to a lesser extent.
This is consistent with the characteristics of the corpus population.
It contains exclusively OOC patients with no laryngeal involvement. 
As treatment effects in these patients primarily affect articulation and prosody, while voice quality is more closely linked to laryngeal function, the lower correlation of voice quality features is expected.
\section{Conclusion}
This work presents an analysis of a Wav2Vec 2.0-based speech intelligibility assessment model, focused on interpretability, using eGeMAPS LLDs as an interpretable reference and the PWCCA-based approach. 
The results show that the model representations are most correlated with Spectral and Prosodic features, while first MFCC coefficient is the most consistently correlated individual LLD across layers. 
These findings align with the known speech impairments of OOC patients (especially in our context where no laryngeal cancer is represented in the corpus), where articulation and prosody are most affected by treatment, reinforcing the clinical trustworthiness of the model. 
Beyond interpretability, this work provides practical guidance on acoustic feature selection. 
Since deep learning models are known for strong performance, features that align closely with their representations represent effective and interpretable alternatives for the community.
Future work should explore other interpretable feature sets beyond eGeMAPS, other SSL architectures, and other pathological speech conditions.
PWCCA combined with a well-established handcrafted feature set represents a promising framework for interpreting deep learning models in clinical speech assessment.
This encourages the community to work towards a unified, clinically grounded feature set that could serve as a standard reference for interpreting deep learning models in speech and voice disorders.

\section{Acknowledgments}
This research was funded, in whole, by Chair LIAvignon, and in part, by the French National Research Agency (ANR), project OLINPIC (ANR-24-CE38-2819). This work was granted access to the HPC resources of IDRIS under the allocation 2025-AD011016558 made by GENCI. For the purpose of open access, the author has applied a CC-BY public copyright license to any Author Accepted Manuscript (AAM) version arising from this submission. 

\section{Generative AI Use Disclosure}
Generative AI tools were used for editing and polishing the manuscript. All scientific content, ideas, analyses, and conclusions are solely the work of the authors.

\bibliographystyle{IEEEtran}
\bibliography{mybib}

\end{document}